\documentclass[aps,showpacs,11pt]{revtex4}
\usepackage{dcolumn}
\usepackage{graphicx}
\usepackage{amsmath}
\usepackage{amsfonts}
\usepackage{amssymb}
\usepackage{psfrag}
\usepackage{wrapfig}
\usepackage{subfigure}
\usepackage{makeidx}
\usepackage{bm}
\usepackage{epsf}
\usepackage{hyperref}
\usepackage{color}
\usepackage{multirow,dcolumn}
\usepackage{graphicx}

\begin{document}

\title{Horizon instability of massless scalar perturbations of an extreme Reissner-Nordstr\"om-AdS black hole}

\author{Shao-Jun Zhang$^1$, Qiyuan Pan$^2$, Bin Wang$^1$, Elcio Abdalla$^3$}
\affiliation{$^1$Department of Physics and Astronomy, Shanghai Jiao Tong University, Shanghai 200240, China\\
$^2$Institute of Physics and Department of Physics,
Hunan Normal University, Changsha, Hunan 410081, China\\
$^3$Instituto de F$\acute{i}$sica, Universidade de S$\tilde{a}$o Paulo, C.P.66.318, CEP 05315-970, S$\tilde{a}$o Paulo, Brazil}

\date{\today}

\begin{abstract}

\indent We study the stability of extreme
Reissner-Nordstr\"om-AdS black hole under
massless scalar perturbations. We show that the
perturbation on the horizon of the extreme
Reissner-Nordstr\"om-AdS black hole experiences a
power-law decay, instead of an exponential decay
as observed in the nonextreme AdS black hole.  On
the horizon of the extreme
Reissner-Nordstr\"om-AdS black hole, the blow up
happens at lower order derivative of the scalar
field compared with that of the extreme
Reissner-Nordstr\"om black hole, which shows that
extreme AdS black holes tend to instability in
comparison to black holes in asymptotic flat
space-times.

\end{abstract}



\maketitle

\section{Introduction}

The black hole, an exotic astronomical object predicted in general relativity, is playing more
and more important roles in various fields of modern physics. Showing the existence of such an
object by studying its stability is obviously important. Starting from the influential study by
Regge and Wheeler \cite{Regge:1957a}, the stability of black holes has been investigated
over half a century. It has been demonstrated that most black holes are stable under various
types of perturbations (for a recent review see for example \cite{Konoplya:2011a}), which shows that the black hole
is realizable in practice and is not just a mathematical curiosity.

Recently, Aretakis carefully examined the stability of the extreme black holes in the
asymptotically flat spacetime. They proved analytically that for the extreme
Reissner-Nordstr\"om (eRN) and Kerr (eKerr) black holes, there exists a classical
instability under massless scalar field perturbations
\cite{Aretakis:2011a,Aretakis:2011b,Aretakis:2011c,Aretakis:2012a}.

This is an amazing result viewed from two aspects. The first one is that non-extreme RN and
Kerr black holes are stable against massless scalar field perturbations
\cite{Blue:2005a,Dafermos:2010a}, so it is surprising that the stability of black holes
changes radically when the extremal limit is approached. The second one is that extreme black
hole occupies an essential place in understanding quantum theory of gravity. For example, the
Bekenstein-Hawking entropy of extremal supersymmetric black holes  can have statistical
explanation in string theory \cite{Strominger:1996a, Guica:2008a}, which reflects a quantum aspect
of gravity. The study on the stability of extreme black holes is obviously important. Aretakis'
result suggests that the stability of extreme black holes has to be reexamined.

Aretakis' argument can be briefly summarized as follows. We work in the ingoing
Eddington-Finkelstein (EF) coordinates $v$ and $r$, and  suppose that the initial data for the
perturbation $\psi$ is defined on a spacelike surface intersecting with the future horizon and
further, one can construct a certain conserved quantity on the horizon, i.e.,
the so-called Aretakis constant. With this constant, it can be proved analytically  that
the derivative $\partial_r \psi$  shall not decay on the horizon. However, the field itself, $\psi$ does decay
on the horizon as well as outside. Moreover, all radial derivatives of $\psi$ fall off outside the horizon. The non-decay of  $\partial_r \psi$ on the horizon leads to a blow up of
$\partial_r^2 \psi$ on the horizon at late times. Aretakis could demonstrate that the derivative
$\partial_r^k\psi$ blows up at the horizon as $v^{k-1}$ or even faster. This
suggests the instability of the extreme black hole. This proof was extended to other extreme
black holes in the asymptotically flat spacetime in various dimensions \cite{Lucietti:2012a}.
Similar instability was also found in the electromagnetic and gravitational perturbations
in the eKerr black hole\cite{Lucietti:2012a}. This type of gravitational instability was also
observed for higher dimensional extreme black holes \cite{Murata:2012a}.

Aretakis's analytic argument was recently confirmed in \cite{Lucietti:2012b} by a numerical
calculation.  They examined the late time behavior of the scalar field in great detail and
found that in the asymptotically flat spacetime the mode of the massless scalar perturbations
with higher $l$ (where $l$ is the spherical harmonic index) decays faster. They also
discussed the case provided that one can not define the non-zero Aretakis constant where
Aretakis' analytical argument breaks down and found that the horizon instability still exists
 (see also the recent work  \cite{Bizon:2012a,Aretakis:2012b} where an analytical argument is given for this case).
Moreover, they showed that for a massive scalar field as well as for electromagnetic and gravitational perturbations, an instability also develops.

In this paper, we will extend the study to the extreme Reissner-Nordstr\"om (eRN-AdS)
black hole. Motivated by the recent discovery of the AdS/CFT correspondence, the investigation of
the stability of AdS black holes becomes more appealing. The stability of the nonextreme AdS
black holes has been studied extensively, see for example \cite{Konoplya:2011a} and references therein. In the
nonextreme AdS black hole, it was shown that the behavior of perturbations differs a lot from
those in the asymptotically flat spacetimes. For example, the scalar field experiences an
exponential decay at late time in the AdS black hole background \cite{Horowitz:1999a,Horowitz:1999b,Wang:2000b}
instead of a power-law decay in the asymptotically flat black holes
\cite{Bicak:1972a,Gundlach:1993a,Gundlach:1993b,Burko:1997a}. Higher $l$ modes will experience an increase
of the damping time scale and a decrease of the oscillation time scale compared with the lower
$l$ modes in the AdS black hole \cite{Wang:2000a,Wang:2000b,Zhu:2001a,Wang:2004a}, while
the situation is opposite in the asymptotically flat black hole \cite{Gundlach:1993a}. These
differences in the behaviors of perturbations are caused by the different effective potentials and
boundary conditions in the AdS black hole compared with the asymptotically flat black hole.
Here we want to examine how the presence of the negative cosmological constant affects the
stability of extreme black holes in the AdS space. We will focus on the massless scalar field
perturbations. In \cite{Wang:2004a}, it was argued that when the RN-AdS black hole approaches
to the extremal limit, the late time decay of the scalar field outside the black hole changes from
exponential to power-law. It is interesting to check this result in the exactly extreme RN-AdS
black hole background both outside the horizon and on the horizon. We will examine the stability
of the extreme RN-AdS black hole and compare the result with that in the asymptotically flat black holes.

This paper is organized as follows. In section II, we will follow Aretakis argument to study the
horizon stability of eRN-AdS black hole analytically. In section III, we will use
numerical method to study the behavior of the massless scalar field perturbations for different
angular index $l=0,1$ and $2$, respectively. Section IV is devoted to conclusion and discussion.
In order to compare with the eRN case, we will set all parameters to be the same as in \cite{Lucietti:2012b}.

\section{Horizon instability: Analytical results}

In this section, we will follow Aretakis' argument
\cite{Aretakis:2011a,Aretakis:2011b,Aretakis:2011c,Aretakis:2012a}
to study the stability of extremal RN-AdS black hole under massless scalar perturbations
analytically. The metric of the extremal RN-AdS black hole takes the form
\begin{eqnarray}
ds^2&=&-f(r)dt^2+f^{-1}(r)dr^2+r^2 d\Omega^2\quad ,\nonumber\\
f(r)&=&\frac{1}{R^2 r^2} \left(r^2+2r_+ r+R^2+3r_+^2\right) (r-r_+)^2 \quad ,\label{metric}
\end{eqnarray}
where $r_+$ denotes the degenerate horizon, and $R$ is the AdS radius related to the cosmological
constant $\Lambda$ by $\Lambda=-3/R^2$. The tortoise coordinate is
\begin{eqnarray}
r_\ast(r)&=&\int \frac{dr}{f(r)}\nonumber\\
&=&\frac{R^2}{\left(R^2+6r_+^2\right)^2} \Bigg[-\frac{r_+^2\left(R^2+6 r_+^2\right)}{r-r_+}+\frac{R^4+7R^2 r_+^2 +14 r_+^4}{\sqrt{R^2+2r_+^2}} \arctan\left(\frac{r+r_+}{\sqrt{R^2+2 r_+^2}}\right) \nonumber\\
&&\qquad +2 r_+ \left(R^2+4 r_+^2\right) \log\left(r-r_+\right)-r_+ \left(R^2+4 r_+^2\right)\log\left(r^2+2r r_+ +3r_+^2+R^2\right)\Bigg].\label{tortoise}
\end{eqnarray}

With the ingoing Eddington-Finkelstein (EF) coordinates, the metric becomes
\begin{eqnarray}
ds^2=-f(r) dv^2+2dv dr +r^2
d\Omega^2,\label{metric1}
\end{eqnarray}
where $v=t+r_\ast$.

The dynamics of the massless scalar perturbation is governed by the Klein-Gordon equation
\begin{eqnarray}
\nabla^2 \psi=0.\label{scalarEoM}
\end{eqnarray}

Working in the EF coordinate (\ref{metric1}), we begin by expanding $\psi$ as
\begin{eqnarray}
\psi (v,r,\Omega)=\sum_{l=0}^{\infty} \psi_l(v,r)
Y_l(\Omega), \label{expansion}
\end{eqnarray}
 where the index $m$ has been dropped. Substituting it into the equation of motion, we
obtain
\begin{eqnarray}
2r \partial_v \partial_r (r\psi_l)+\partial_r (\Delta \partial_r
\psi_l) -l(l+1) \psi_l=0,\label{scalarEoM1}
\end{eqnarray}
where $\Delta=\frac{1}{R^2} \left(r^2+2r_+ r+R^2+3r_+^2\right) (r-r_+)^2$. We work with zero angular momentum ($l=0$) and compute the expression  at $r=r_+$. We can thus show that
\begin{eqnarray}
H_0[\psi]\equiv \frac{1}{r_+} \left[\partial_r (r\psi_0)\right]_{r=r_+},\label{H0}
\end{eqnarray}
is independent of $v$.

The constant $H_0 $ does not vanish for general initial data, thus remaining non-zero.
As a consequence the field and its radial derivative do not simultaneously tend to zero at the
horizon. Later, we show how $\psi$ decays
at late times using a numerical computation. Therefore, $\partial_r \psi$ does not decay at the horizon,
\begin{eqnarray}
(\partial_r \psi_0)_{r=r_+} \rightarrow H_0 \qquad {\textrm as} \qquad v\rightarrow \infty\quad .\label{H0-1}
\end{eqnarray}

Now, acting on (\ref{scalarEoM1}) with $\partial_r$ for $l=0$ and $r=r_+$, we obtain
\begin{eqnarray}
\left[\partial_v\partial_r^2 (r\psi_0) +\left(\frac{6}{R^2}+\frac{1}{r_+}\right) \partial_r \psi_0\right]_{r=r_+}=0\quad .\label{scalarEoM2}
\end{eqnarray}
Hence,
\begin{eqnarray}
\left[\partial_v\partial_r^2 (r\psi_0)\right]_{r=r_+} \rightarrow -\left(\frac{6}{R^2}+\frac{1}{r_+}\right) H_0 \qquad {\textrm as} \qquad v\rightarrow \infty\quad .\label{H0-2}
\end{eqnarray}
This result leads to the fact that the second radial derivative of the field diverges for large $v$ at the horizon, (similar conclusion being true for higher derivatives, as one can easily see deriving
(\ref{scalarEoM1}))
\begin{eqnarray}
\left(\partial_r^2 \psi_0\right)_{r=r_+} \sim -\left(\frac{6}{r_+ R^2}+\frac{1}{r_+^2}\right) H_0 v \qquad {\textrm as} \qquad v\rightarrow \infty\quad .\label{H0-3}
\end{eqnarray}

Such instability is typical for  $l=0$
perturbations. For $l>0$, because
of the complexity of $\Delta$ in
(\ref{scalarEoM1}), we can not define a conserved
Aretakis constant by taking $\partial_r^l$ on the
equation of motion (\ref{scalarEoM1}) and
evaluating it at the horizon as in the eRN
case \cite{Lucietti:2012b}. Thus, in contrast with the
eRN case \cite{Lucietti:2012b}, the above
analytic analysis cannot be extended to arbitrary
$l>0$.

However, for $l>0$ modes, we can still get some
important information by analyzing the
asymptotical property of the equation of motion
(\ref{scalarEoM1}) at large $v$ by supposing that
the late-time behavior of $\phi\equiv r\psi_l$ on
the horizon is about $v^{a_0}$ with $a_0$ a
negative non-integer constant. We will prove this
assumption and get $a_0$ later using numerical
method. Then from (\ref{scalarEoM1}), at large
$v$, we get that $\partial_r \phi |_{r=r_+} \sim
v^{a_0+1}$. By taking further $r$-derivatives of
(\ref{scalarEoM1}), and analyzing the behavior at
large $v$, we can obtain
\begin{eqnarray}
\partial_r^n \phi |_{r=r_+} \sim v^{a_0+n} \qquad {\textrm as} \qquad v\rightarrow\infty.\label{latetime}
\end{eqnarray}
This implies that there is always horizon instability for large enough $n$.

\section{Numerical results for massless scalar perturbations}

In this section, we use a numerical method to analyze the behavior of the massless scalar
perturbations along the horizon in the background of the eRN-AdS black hole. In the analytic
analysis, we required the general initial condition to be of the form of nonzero Aretakis
constant $H_0[\psi]$. This is equivalent to consider an initial outgoing wavepacket in the
perturbation \cite{Lucietti:2012b}. Besides the outgoing wavepacket,  we can also have an initial
ingoing wavepacket, which corresponds to zero Aretakis constant where the analytical argument
above does not work. In this section, we will examine the behavior of massless scalar perturbations
in the eRN-AdS background carefully by imposing both the outgoing and ingoing initial wavepackets.

\subsection{Double null coordinates}

With $(u,v)$-coordinates defined as
\begin{eqnarray}
du=dt-dr_\ast, \quad dv=dt+dr_\ast\quad ,\label{uv}
\end{eqnarray}
the metric (\ref{metric1}) becomes
\begin{eqnarray}
ds^2=-f(r(u,v)) du dv+r^2 d\Omega\quad .\label{uv_metric}
\end{eqnarray}
The areal radius $r(u,v)$ can be determined by solving the tortoise coordinate $r_\ast(r)=\frac{v-u}{2}$.

We wish a nonsingular metric at the horizon, leading us to define new coordinates as in \cite{Lucietti:2012b},
that is,
\begin{eqnarray}
\frac{u}{2}&=&-r_\ast(r_+-U)\nonumber\\
&=&-\frac{R^2}{\left(R^2+6r_+^2\right)^2} \Bigg[\frac{r_+^2\left(R^2+6 r_+^2\right)}{U}+\frac{R^4+7R^2 r_+^2 +14 r_+^4}{\sqrt{R^2+2r_+^2}} \arctan\left(\frac{2r_+-U}{\sqrt{R^2+2 r_+^2}}\right) \nonumber\\
&&\qquad +2 r_+ \left(R^2+4 r_+^2\right) \log\left(-U\right)-r_+ \left(R^2+4 r_+^2\right)\log\left(U^2- 4 r_+ U +6r_+^2+R^2\right)\Bigg].\label{U}
\end{eqnarray}

In $(U,v)$-coordinates, the position of the horizon is  at $U=0$ and in the region $U<0$ one is
outside the black hole. The metric becomes
\begin{eqnarray}
ds^2=-\frac{2f(r)}{f(r_+-U)} dU dv+r^2 d\Omega^2,\label{Uv_metric}
\end{eqnarray}
where $r$ is a function of $U$ and $v$. We can expand $r$ for small $U$ as
\begin{eqnarray}
r=r_+ -U+\frac{1}{2} \left(\frac{1}{r_+^2}+\frac{6}{R^2}\right) U^2 +\left[\left(\frac{1}{r_+^3}+\frac{4}{R^2 r_+}\right)v - \left(\frac{9}{R^4}+\frac{1}{4r_+^4}+\frac{3}{R^2 r_+^2}\right) v^2\right] U^3+ \cdots\quad , \label{r_expansion}
\end{eqnarray}
from which it can be shown that $\frac{f(r)}{f(r_+-U)}=1+{\cal O}(U)$ for small
$U$. We obtain a regular analytic metric that can be defined for $U>0$.

\subsection{Wave equation and initial data}

Defining $\phi\equiv r\psi_l$, where $l$ is the angular index, we obtain a wave equation for
$\phi$ in $(U,v)$-coordinates from the Klein-Gordon equation (\ref{scalarEoM})
\begin{eqnarray}
4 \partial_U \partial_v \phi+\hat{V} (U,v) \phi=0\quad ,\label{wave-equation}
\end{eqnarray}
where the effective potential
\begin{eqnarray}
\hat{V} (U,v)=\frac{2 f(r)}{f(r_+-U)} \left(\frac{f'(r)}{r} +
\frac{l(l+1)}{r^2}\right).
\end{eqnarray}

 We consider a null ``initial" surface as in \cite{Lucietti:2012b}
\begin{eqnarray}
\Sigma_0 = \{U=U_0, v\geq v_0\} \cup \{U\geq U_0, v=v_0\}\quad ,\label{initial-surface}
\end{eqnarray}
and  impose the following two types of initial data:

\begin{itemize}

\item outgoing wavepacket
\begin{eqnarray}
\phi(U,v_0)=\exp\left(-\frac{(U-\mu)^2}{2 \sigma^2}\right),\quad \phi(U_0,v)=0.\label{outgoing}
\end{eqnarray}

\item ingoing wavepacket
\begin{eqnarray}
\phi(U,v_0)=0, \quad \phi(U_0,v)=\exp\left(-\frac{(v-\mu')^2}{2 \sigma'^2}\right).\label{ingoing}
\end{eqnarray}

\end{itemize}

We will solve the perturbation equation numerically by using the above initial conditions.
In order to do the comparison with the eRN results \cite{Lucietti:2012b}, we also set
$r_+=1, R=1$, $U_0=-0.5$ and $v_0=0$ in the numerical computation.

\subsection{Algorithm of numerical method}

We apply the finite difference method suggested in \cite{Gundlach:1993a,Gomez:1992a} to solve the
wave equation (\ref{wave-equation}), which can be discretized into
\begin{eqnarray}
\phi_N=\phi_E+\phi_W-\phi_S-\delta U \delta v \hat{V}\left(\frac{v_N+v_W-u_N-u_E}{4}\right) \frac{\phi_W+\phi_E}{8} +{\cal O}(\epsilon^4)\quad ,\label{discretization}
\end{eqnarray}
where points $N, S, E$ and $W$ form a null rectangle with relative positions as: $N:
(U+\delta U, v+\delta v), W: (U+\delta U, v), E: (U, v+\delta v)$ and $S: (U, v)$. The parameter
$\epsilon$ is an overall grid scalar factor, so that $\delta U\sim \delta v \sim \epsilon$.

There is one essential point we should note: the effective potential $\hat{V}$ is positive and
vanishes at the horizon, but it diverges at $r\rightarrow \infty$, which requires that $\phi$
vanishes at the infinity. This is the boundary condition to be satisfied by the wave equation
for the scalar field in the AdS space, which is completely different from that in the
asymptotically flat space. In the perturbations of the nonextreme AdS black hole, it is this
difference that makes the perturbation behave differently from that in the asymptotically flat
spacetime. In the following, we will examine the effective potential effect in the perturbation in
the eRN-AdS black hole and compare with that in the asymptotically flat spacetimes. In terms of
the tortoise coordinate $r_\ast$, it is seen that when $r$ tends to infinity, $r_\ast$ tends to a
finite constant, which is denoted as $r_\ast^{as}$. It means that our region of interest in the
$(U-v)$ diagram is below the curve $v-u(U)=2 r_\ast^{as}$, as shown in figure (1). On  this line we
set $\phi=0$, since there $r\rightarrow \infty$ and the effective potential diverges.

The inversion of the relation $r_\ast (r)$ needed in the evaluation of the potential $\hat{V}(U,v)$
is the most tedious part in the computation. We overcome this difficulty by employing the method
suggested in \cite{Gundlach:1993a,Brady:1999a}.

\begin{center}

\begin{figure}[!htbp]
\includegraphics[bb=188 314 454 511]{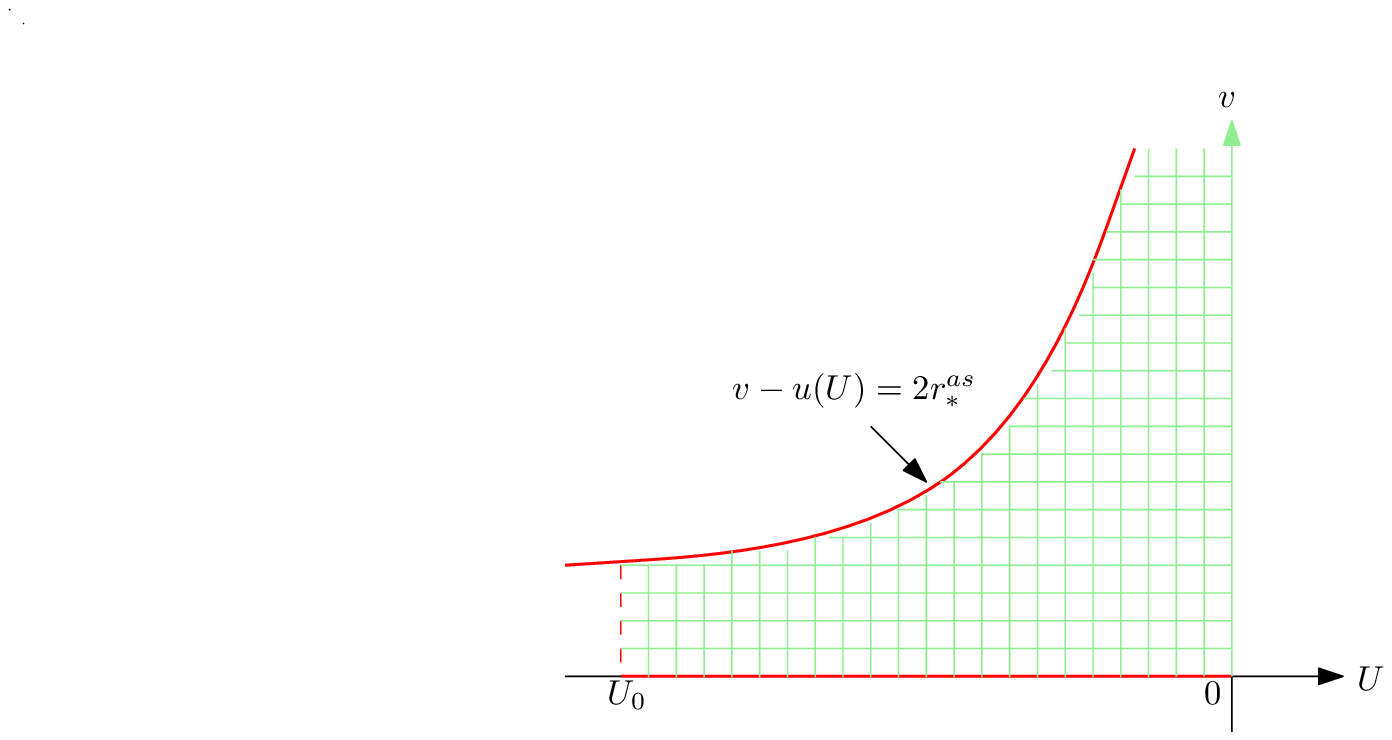}
\caption{Sketch graph of our interested region in
$(U,v)$-coordinates filled with light-green mesh. Values of the
scalar field on the three red lines, $v-u(U)=2r_\ast^{as}, U=U_0$
and $v=0$, are known according to the given initial data. In this
paper, we are most interested in values of the scalar field on the
horizon, which is the line $U=0$ we show with light-green color.}
\end{figure}
\end{center}

\subsection{Numerical results}

\subsubsection{The $l=0$ mode}

Heret we report on the numerical result of solving the wave equation (\ref{wave-equation})
with $l=0$. We define Aretakis' conserved quantity as in Eq. (\ref{H0}). The outgoing wave
initial data (\ref{outgoing}) has nonzero $H_0[\psi]$ unless $\mu=0$, while the ingoing
wave initial data (\ref{ingoing}) and outgoing wave initial data (\ref{outgoing}) with $\mu=0$
have zero $H_0[\psi]$. We present the results of numerical computations by using different initial
conditions in the following. To  make a comparison with results in eRN case \cite{Lucietti:2012b},
we choose the same parameters for the perturbations.

 \textit{1.1 Non-zero Aretakis constant}

Firstl, we consider the solution with $H_0[\psi]
\neq 0$, where we use the initial outgoing
wavepacket (\ref{outgoing}) with $\mu \neq 0$. We
have shown the  instability of this type of
perturbation analytically in the last section,
where we assumed that $\psi$ decays on the
horizon. Here we will show that this assumption
holds through numerical computation.

As in ref. \cite{Lucietti:2012b}, although we do
our numerical calculations in $(U,v)$
coordinates, we would like to display the results
using $(v,r)$ coordinates, since they correspond to the
preferred  coordinates, related to the
symmetries of the background.

\begin{figure}[!htbp]
\centering
\includegraphics[scale=0.4]{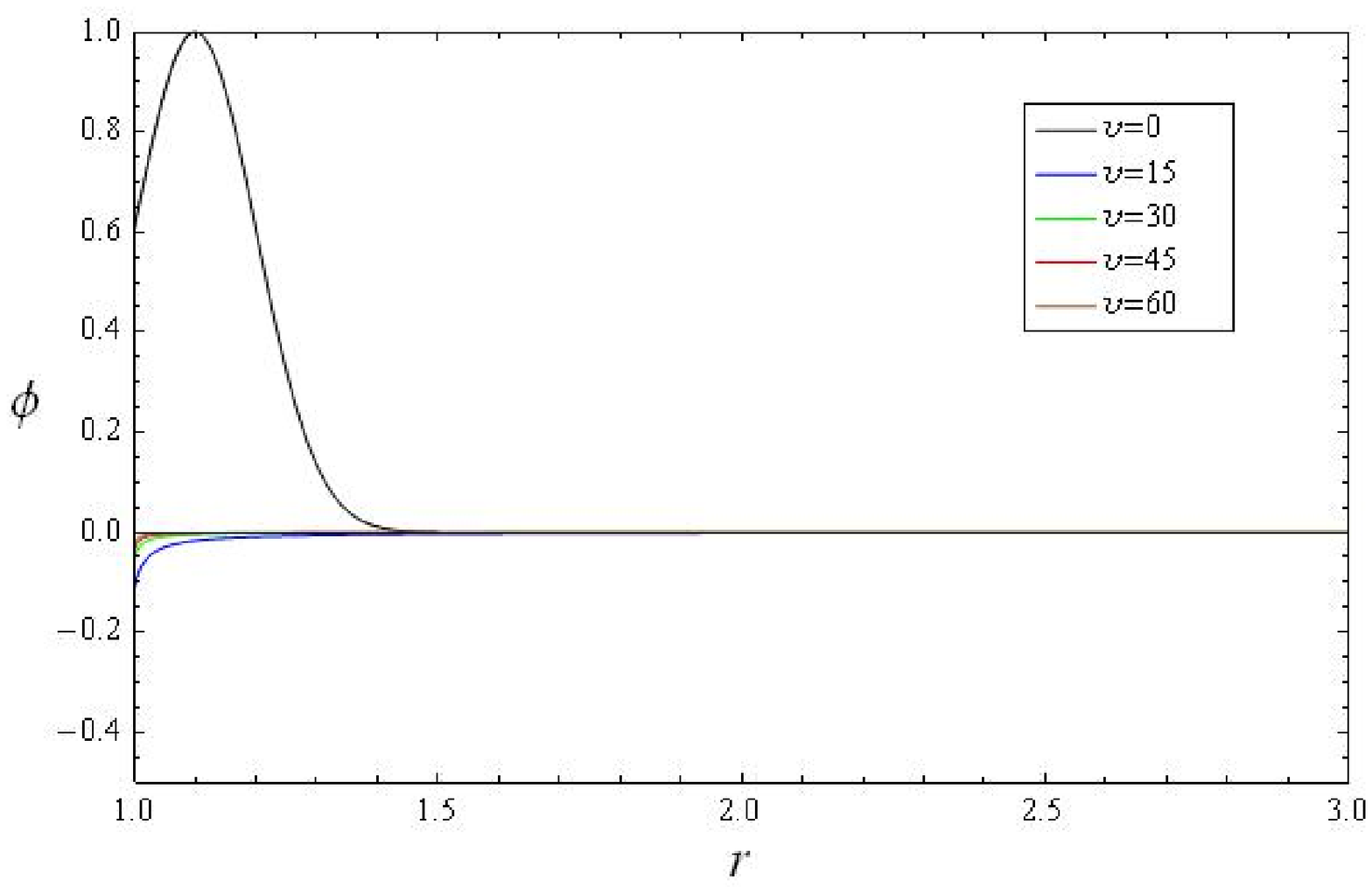}
\includegraphics[scale=0.4]{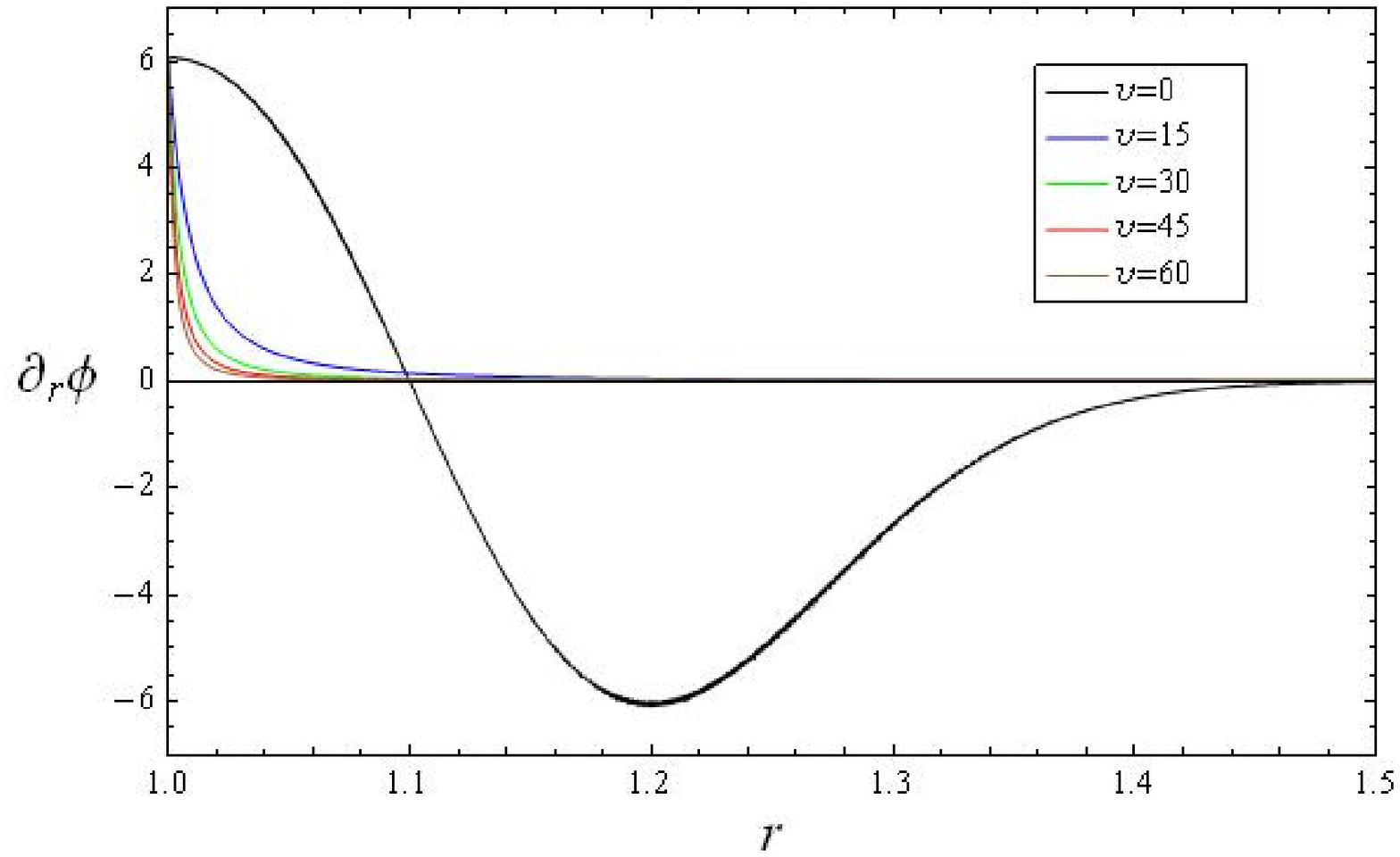}
\caption{Functions $\phi(v,r)$ and $\partial_r \phi(v,r)$ for $l=0$ on fixed $v$ slices. Seeking comparison with other results, the outgoing wave initial data is given by (\ref{outgoing}) with $(\sigma, \mu)=(0.1,-0.1)$. As we can see, $\phi$ decays on and outside the horizon. And $\partial_r \phi$ also decays outside the horizon, while keeping constant on the horizon. It is  steeper near the horizon when $v$ increases. Therefore, the second derivative of  $\phi$ diverges at the horizon.}
\end{figure}

In Fig. 2, the time evolution of $\phi$ and
$\partial_r \phi$ in $(v,r)$-coordinate is
plotted with $(\sigma, \mu)=(0.1,-0.1)$. The
figure shares some similar features  in the eRN
case in \cite{Lucietti:2012b}: (1)  as $v$ increases, $\phi$ decays;
(2) $\partial_r \phi$ also
decays outside the horizon, but does not decay on
the horizon; (3) Moreover, the first derivative of  $\phi$
becomes steeper next to the horizon as $v$
increases, indicating that $\partial_r^2
\phi$ must blow up along the horizon. Besides the
similarity,  we also observe the differences
compared with eRN case. $\phi$ and $\partial_r
\phi$ outside the horizon decay faster in eRN-AdS
case. $\partial_r \phi$ becomes steeper more
quickly near the horizon as $v$ increases in the
eRN-AdS black hole, which shows that
$\partial_r^2 \phi$ on the horizon blows up more
violently in the AdS background.

\begin{figure}[!htbp]
\centering
\includegraphics[scale=0.48]{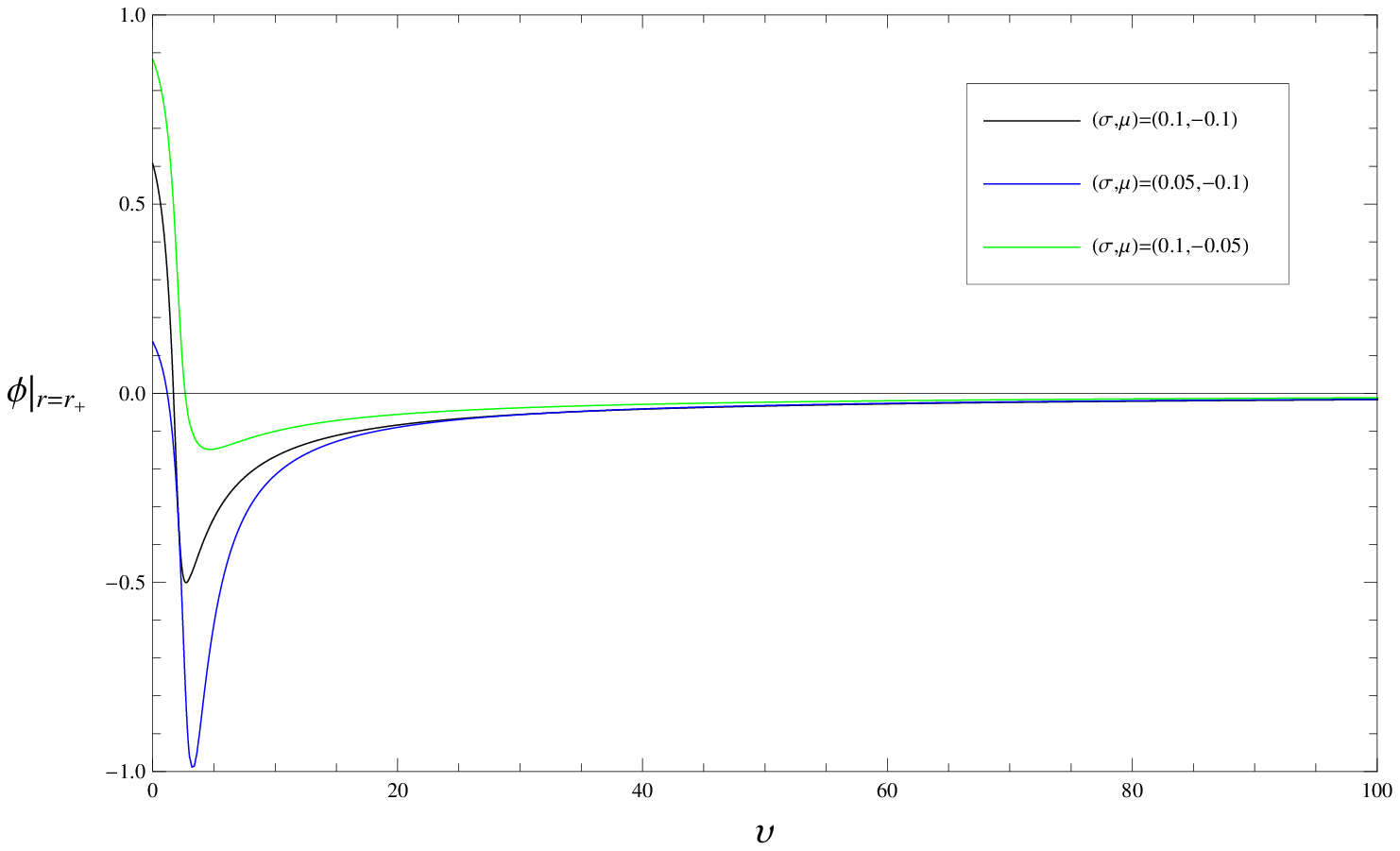}
\includegraphics[scale=0.55]{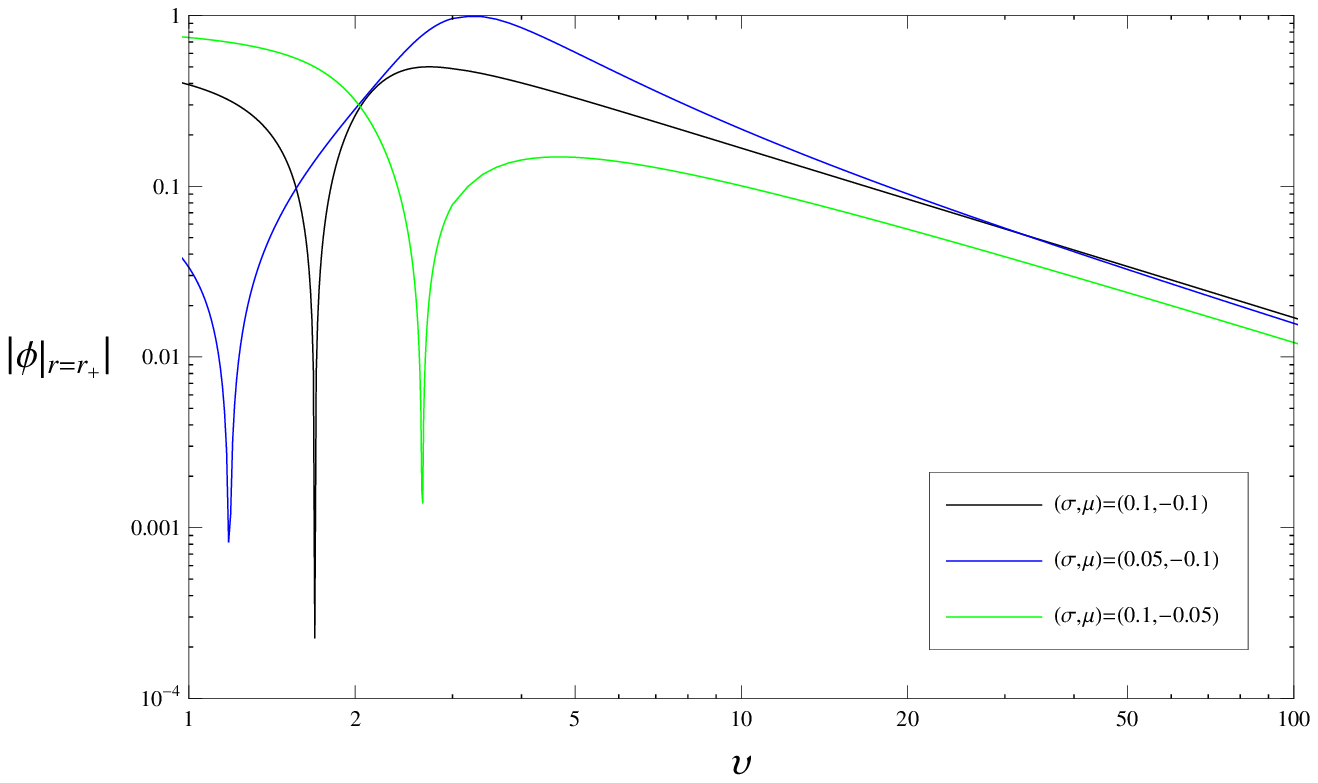}
\includegraphics[scale=0.55]{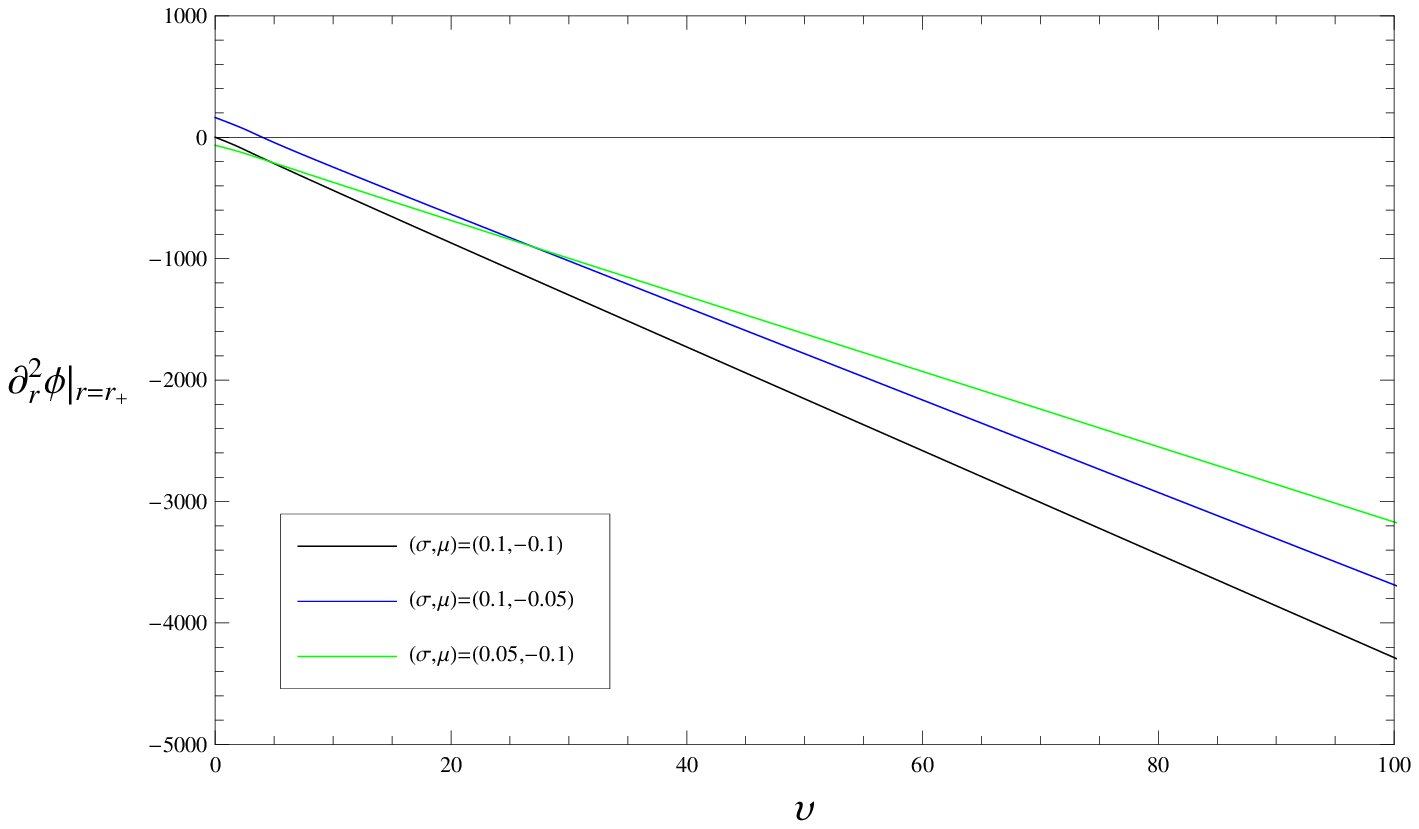}
\caption{Time evolution of $\phi|_{r=r_+}$ and
$\partial_r^2 \phi |_{r=r_+}$ for $l=0$ mode. We
consider outgoing wave initial data
(\ref{outgoing}) with various parameters:
$(\sigma,\mu)=(0.1, -0.1), (0.05, -0.1), (0.1,
-0.05)$. From the top-right Log-Log plot, we can
see that $\phi |_{r=r_+}$ has a power-law decay.
While $\partial_r^2 \phi |_{r=r_+}$ blows up as
$v$ increases, which can be seen from the bottom
plot.}
\end{figure}

In Fig. 3, we have the time evolution of the field and its second derivative at the horizon.
We take the constants  $(\sigma,\mu)$ as being $(0.1,
-0.1), (0.05, -0.1), (0.1, -0.05)$ for the sake of comparison. We can see
that $\phi |_{r=r_+}$ can quickly get rid of the
influence of different initial parameters and
exhibit the consistent late time behavior earlier
in the eRN-AdS black hole if we compare with
Fig.4 of the eRN black hole case in
\cite{Lucietti:2012b}. By fitting the absolute
value of $\phi |_{r=r_+}$ in the range $80\leq v
\leq 100$ to $v^a$, we obtain the following
exponents: $a=-0.999, -1.039, -0.980$ for
$(\sigma,\mu)=(0.1, -0.1), (0.05, -0.1), (0.1,
-0.05)$, respectively. These results suggest that
the scalar field on the horizon decays as
$v^{-1}$ at late time, which is the same as that
in the eRN case \cite{Lucietti:2012b}. The
power-law decay in the late-time behavior of the
massless scalar perturbation in eRN-AdS black
hole is very different from results in
non-extremal AdS black holes
\cite{Horowitz:1999a,Horowitz:1999b,Wang:2000b},
where the late time tail exhibit exponential
decay. Our result supports the argument in
\cite{Wang:2004a}, where it was argued that when
AdS black holes approach extremal limit, there is
a transition from exponential decay to power-law
decay. The difference in the late time
perturbation indicates the dynamical difference
between the extreme black hole and nonextreme
black hole in AdS spacetimes. The coefficient of
the power law decay can be fitted to be  $a \sim
-0.28$ for all initial data we used instead of
$a\sim -2$ in the eRN case. Hence, the late time
behavior of $\phi |_{r=r_+}$ for the eRN-AdS
black hole is
\begin{eqnarray}
\phi |_{r=r_+} \sim -\frac{0.28 H_0}{v} \qquad
v\rightarrow \infty\quad .\label{outgoing-late-time}
\end{eqnarray}
$\partial_r^2 \phi |_{r=r_+}$ blows up linearly
as expected in the analytical study. We fit the
curves of $\partial_r^2 \phi |_{r=r_+}$ to a
function $c H_0 v+d$ in the range $80\leq v \leq
100$, and find the fitting parameter $c=-7.070,
-7.022, -7.021$ for $(\sigma,\mu)=(0.1, -0.1),
(0.05, -0.1), (0.1, -0.05)$, respectively. This
suggests that $\partial_r^2 \phi |_{r=r_+} \sim
-7 H_0 v$ at late time, which is consistent with
the analytical result (\ref{H0-3}). For the eRN
case \cite{Lucietti:2012b}, we can do the same
fitting and find $c \sim -1$ for all the initial
data. So we can see that $\partial_r^2 \phi
|_{r=r_+}$ blows up faster in AdS case, which
means that the horizon is more unstable in the
AdS case.

\textit{1.2 Zero Aretakis constant}

Now we consider perturbations with $H_0[\psi]=0$.
The case $\mu =0$ for  an ingoing or an outgoing wavepacket is here contained.
The analytic proof of the instability in the last
section does not work now; instead we will use a
numerical calculation to study this case. To
compare with the eRN case \cite{Lucietti:2012b},
we choose the same parameter spaces $(\sigma',
\mu')=(3.0,10.0)$ for the ingoing wavepacket
(\ref{ingoing}), and $\sigma=0.05, 0.1, 0.15$ and
$\mu=0$ for the outgoing wavepacket
(\ref{outgoing}).

\begin{figure}[!htbp]
\centering
\includegraphics[scale=0.58]{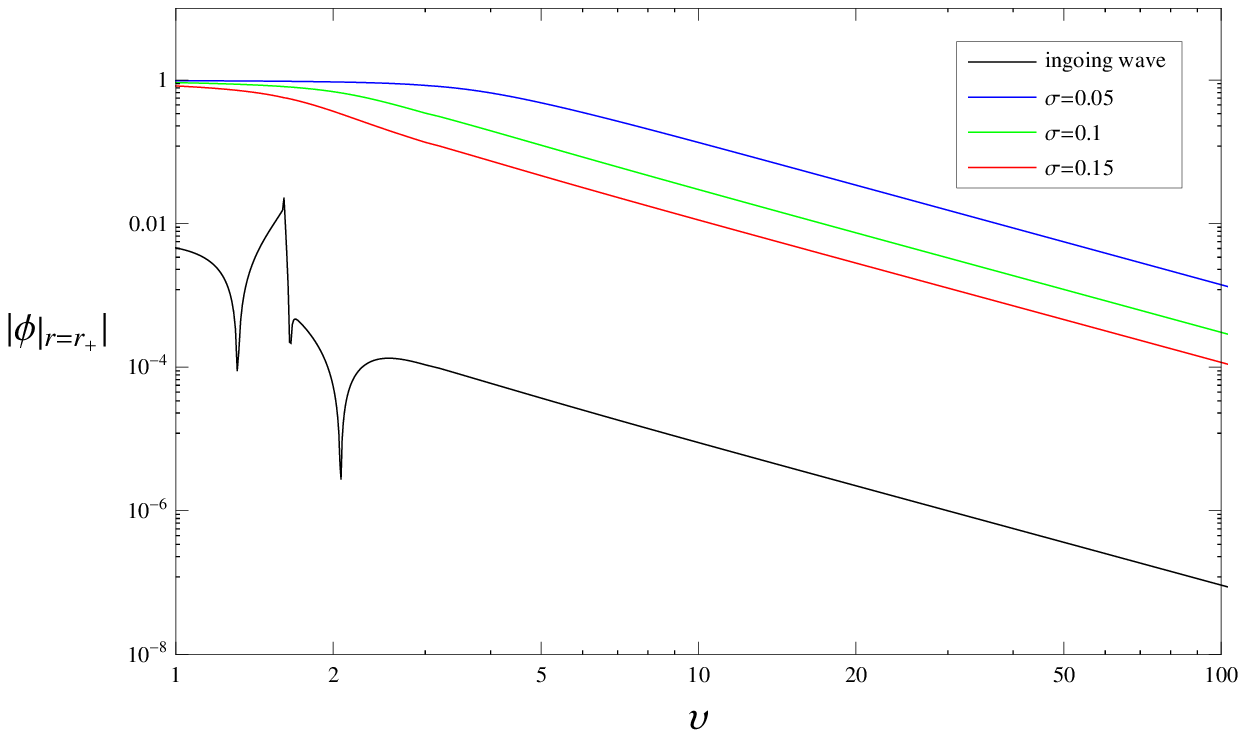}
\includegraphics[scale=0.55]{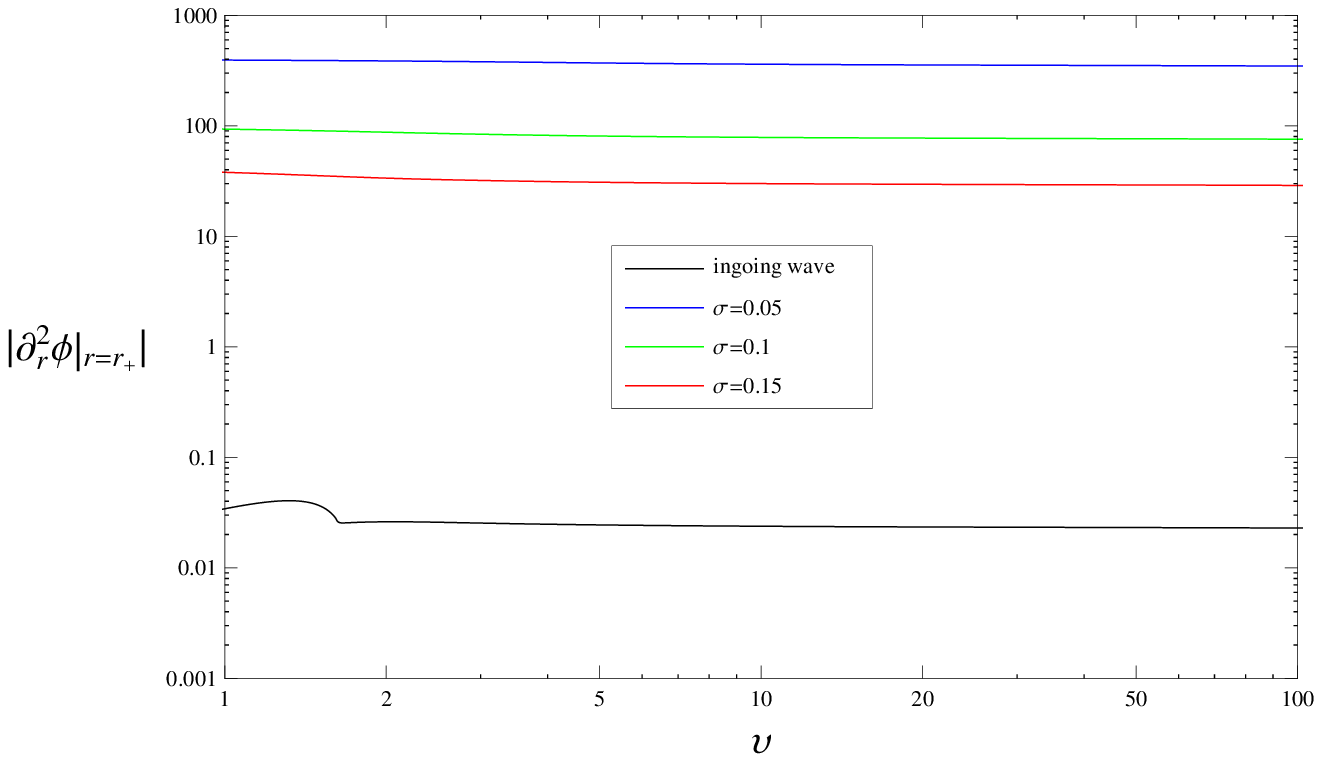}
\includegraphics[scale=0.6]{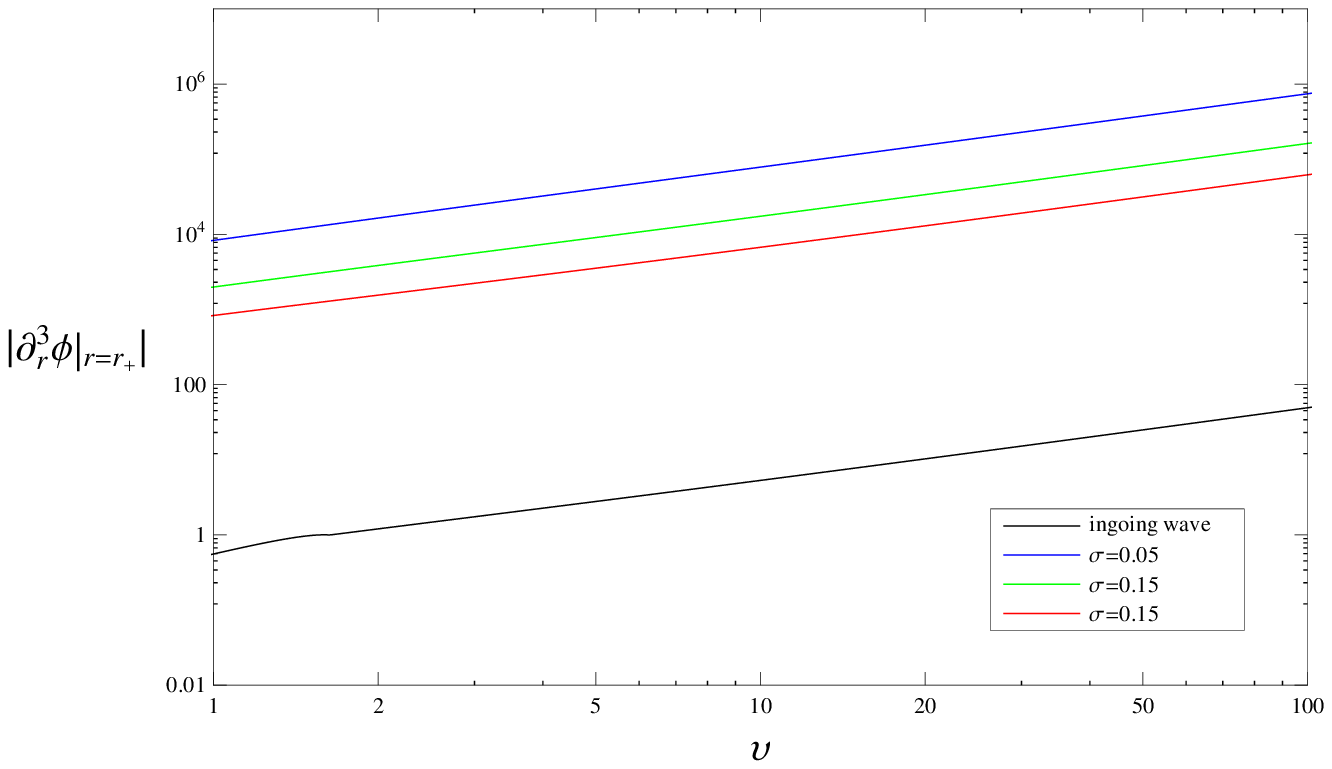}
\caption{Log-Log plots of time evolution of $\phi |_{r=r_+},
\partial_r^2 \phi |_{r=r_+}$ and $\partial_r^3 \phi|_{r=r_+}$ for
$l=0$ and $H_0=0$. The initial data are:
$\sigma=0.05, 0.1, 0.15$ with $\mu=0$ and for
outgoing wavepacket, while  $(\sigma', \mu')=(3.0,
10.0)$ for an ingoing wavepacket. As in previous results, for large values of $v$
the field at the horizon  decays
as $v^{-2}$,  its second derivative at the horizon
approaches a non-vanishing constant, while
the third derivative diverges.  }
\end{figure}

In Fig. 4, we plot the time evolutions of $\phi
|_{r=r_+}, \partial_r^2 \phi |_{r=r_+}$ and
$\partial_r^3 \phi |_{r=r_+}$. We see that $\phi
|_{r=r_+}$ decays, $\partial_r^2 \phi |_{r=r_+}$
approaches to a non-zero constant and
$\partial_r^3 \phi |_{r=r_+}$ blows up as $v$
increases. These behaviors are similar to that
observed in the eRN case in
\cite{Lucietti:2012b}. We see that  there exists
the instability even for initial data with
$H_0[\psi]=0$ in the eRN-AdS black hole.

Using the fitting method, we find that $\phi |_{r=r_+}$
has a power-law decay for the extreme AdS black
hole as argued in \cite{Wang:2004a}, rather than
an exponential decay as we observed in the
nonextreme AdS hole. Fitting to the decay law
$v^a$ at late time, we find the exponent
$a=-1.983$ for the ingoing wave, and $a=-1.989,
-1.986, -1.984$ for the outgoing wave with
$\sigma=0.05, 0.1, 0.15$. This implies that, for
$H_0[\psi]=0$, the late time behavior of $\phi
|_{r=r_+}$ is
\begin{eqnarray}
\phi |_{r=r_+} \sim \frac{C}{v^2} \qquad
v\rightarrow \infty\quad .\label{ingoing_late_time}
\end{eqnarray}
This result is the same as in the eRN case
\cite{Lucietti:2012b}.

Now let us take a closer look at the instability,
which is shown in the late time behavior of
$\partial_r^3 \phi |_{r=r_+}$. By fitting values
of $|\partial_r^3 \phi |_{r=r_+}|$ to function
$v^a$ at late time, we find the fitting parameter
$a=0.979$ for the ingoing wave, and $a=0.982,
0.980, 0.978$ for the outgoing wave with
$\sigma=0.05, 0.1, 0.15$ respectively.  This
confirms that $\partial_r^3 \phi |_{r=r_+}$
indeed blows up as $v$ increases. To determine
the coefficient of the linear blow up, we fit
$|\partial_r^3 \phi |_{r=r_+}|$ to the function
$b v+c$ for $50\leq v\leq 100$, where we find
$b=0.483$ for the ingoing wave and $b=7322.14,
1593.8, 608.642$ for the outgoing wave with
$\sigma=0.05, 0.1, 0.15$ respectively. We can
also do the same fitting for the results in eRN
case \cite{Lucietti:2012b}, which gives $b=3.198$
for the ingoing wave, and $b=921.548, 190.923,
65.965$ for the outgoing wave with $\sigma=0.05,
0.1, 0.15$ respectively.  Comparing with the eRN
case, we observe that the instability is moderate
for the ingoing perturbation and more violent for
the outgoing perturbation in the eRN-AdS case.

\subsubsection{The $l=1$ mode}

For $l=1$ mode, we can no longer define an
Aretakis constant $H_1$ as did in the eRN case
\cite{Lucietti:2012b}. So we can not
classify perturbations according to whether $H_1$
is zero or not. But for convenience to do the
comparison with the eRN case, we will still
classify perturbations into two classes, Type I
and Type II perturbations, according to the
outgoing and ingoing initial wavepackets we
choose.

\textit{2.1 Type I perturbations}

 In this part, we consider type I perturbations, an outgoing
wavepacket (\ref{outgoing}) with  the same
parameters as chosen in \cite{Lucietti:2012b}:
$(\sigma, \mu)=(0.1, 0), (0.1, -0.05), (0.05,
0)$. These correspond to perturbations with
non-zero Aretakis constant in eRN case.

\begin{figure}[!htbp]
\centering
\includegraphics[scale=0.46]{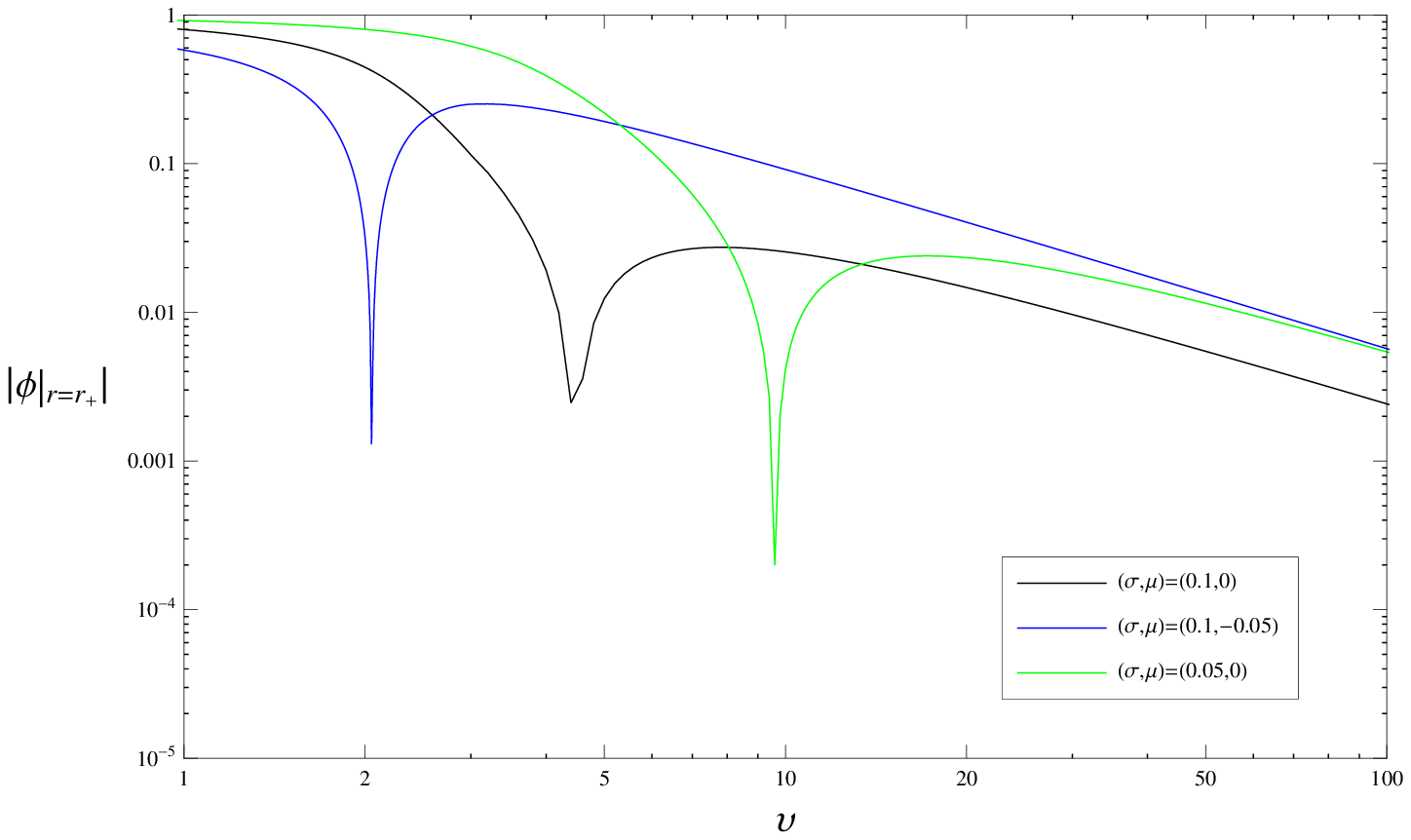}
\includegraphics[scale=0.53]{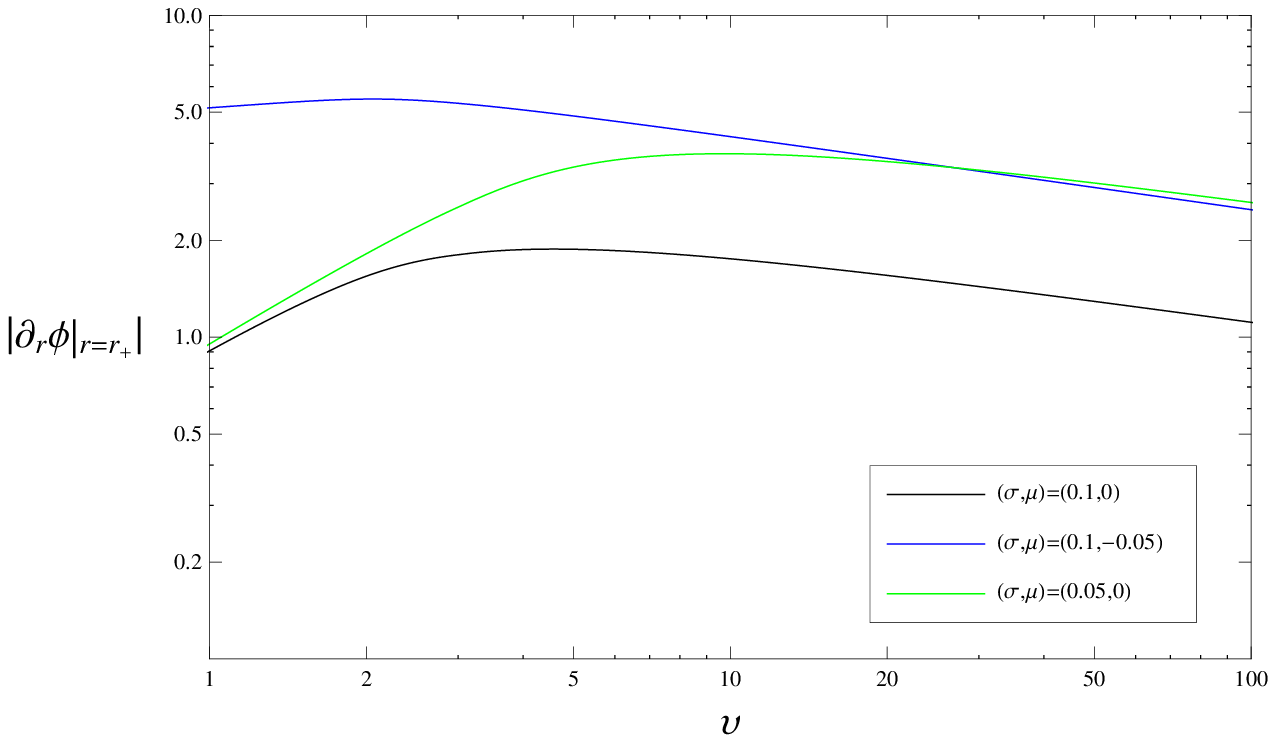}
\includegraphics[scale=0.53]{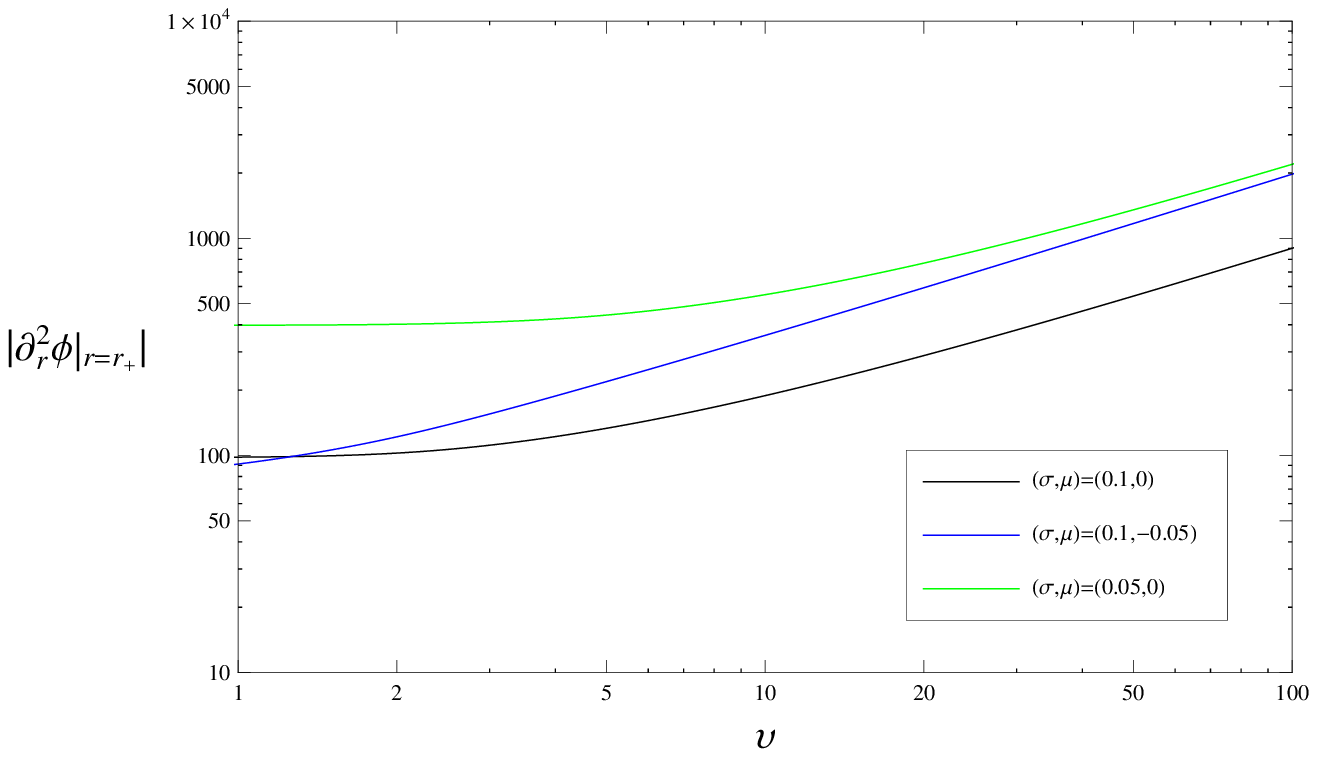}
\includegraphics[scale=0.55]{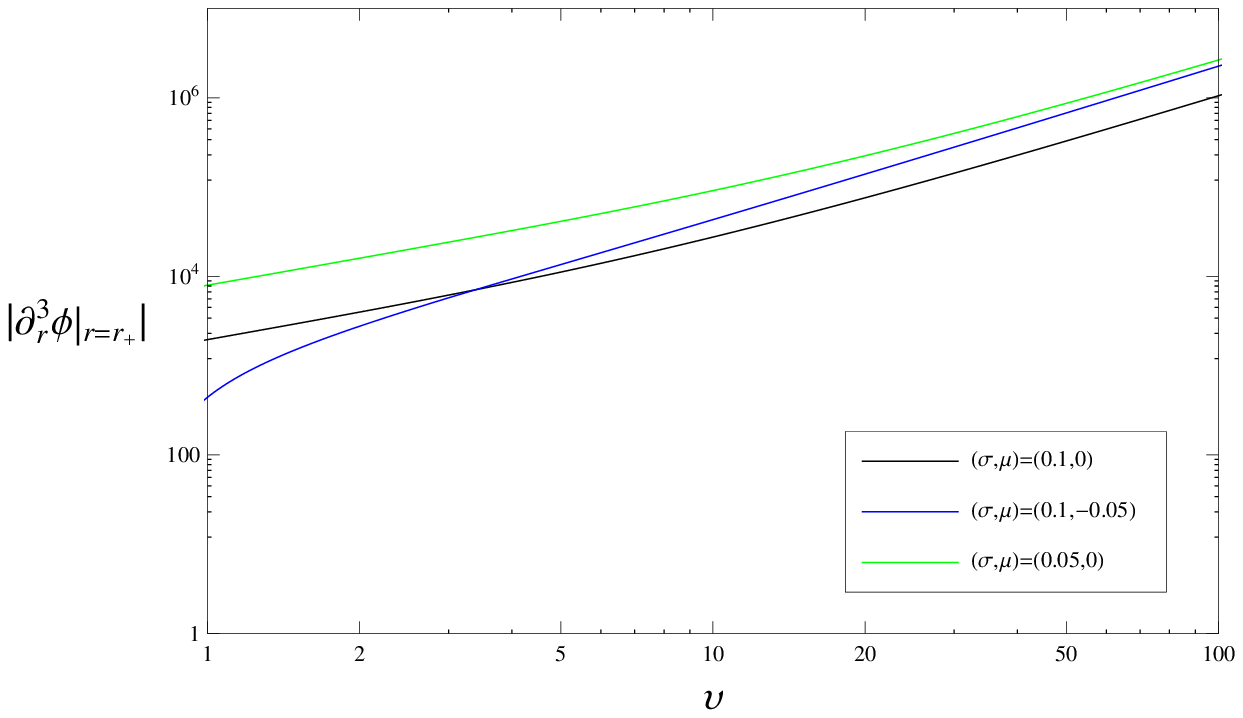}
\caption{Time evolution of $\phi |_{r=r_+},
\partial_r \phi |_{r=r_+}, \partial_r^2 \phi
|_{r=r_+}$ and $\partial_r^3 \phi |_{r=r_+}$ for
$l=1$.  We use the initial data as
$(\sigma, \mu)=(0.1, 0), (0.1, -0.05), (0.05,
0)$. We can see that: $\phi |_{r=r_+}$ and
$\partial_r \phi |_{r=r_+}$ exhibit a power-law
decay, while $\partial_r^2 \phi |_{r=r_+}$ and
$\partial_r^3 \phi |_{r=r_+}$ take a power-law
blow-up. This implies that, for $l=1$ mode,
horizon instability starts to appear from the
second derivative of $\phi$.}
\end{figure}

In Fig. 5, we plot the time evolution of $\phi
|_{r=r_+}, \partial_r \phi |_{r=r_+},
\partial_r^2 \phi |_{r=r_+}$ and $\partial_r^3
\phi |_{r=r_+}$. $\phi |_{r=r_+}$ and $\partial_r
\phi |_{r=r_+}$ are shown in the first two plots
in Fig. 5. They both exhibit power-law decays,
but $\partial_r \phi |_{r=r_+}$ has a slower
decay. Fitting the absolute value of $\phi
|_{r=r_+}$ to the function $v^a$ at late time, we
find the fitting exponent $a=-1.191, -1.230,
-1.127$ for $(\sigma, \mu)=(0.1, 0), (0.1,
-0.05), (0.05, 0)$ respectively. This suggests
that the late time behavior of $\phi |_{r=r_+}$
is
\begin{eqnarray}
\phi |_{r=r_+} \sim C v^{-6/5} \qquad v\rightarrow \infty\quad , \label{L1_outgoing-decay}
\end{eqnarray}
which is very different from the result in eRN
case \cite{Lucietti:2012b}, where $\phi |_{r=r_+}
\sim v^{-2}$ as $v\rightarrow \infty$. Fitting
the absolute value of $\partial_r \phi |_{r=r_+}$
to $v^a$ at late time, we find  $a=-0.218,
-0.229, -0.206$ for $(\sigma, \mu)=(0.1, 0),
(0.1, -0.05), (0.05, 0)$ respectively.

Now we investigate the instability. We can see
that the instability starts to appear in the
second derivative of $\phi$ on the horizon. While
in eRN case, the horizon instability starts to
appear in the third derivative. By fitting the
value of $|\partial_r^2 \phi |_{r=r_+}|$ to $v^a$
at late time, we find the fitting parameter
$a=0.740, 0.757, 0.713$ for $(\sigma, \mu)=(0.1,
0), (0.1, -0.05), (0.05, 0)$ respectively.
Fitting the value of $|\partial_r^3 \phi
|_{r=r_+}|$ to $v^a$ for $80 \leq v \leq 100$, we
find  $a=1.703, 1.746, 1.645$ for $(\sigma,
\mu)=(0.1, 0), (0.1, -0.05), (0.05, 0)$
respectively.

For clarity, we list all the fitting results
above in Table I.

 \textit{2.2 Type II pertubations}

Now we consider type II perturbations, where we
take an ingoing wavepacket (\ref{ingoing}) with
$(\sigma', \mu')=(3.0, 10.0)$ or an outgoing
wavepacket (\ref{outgoing}) with  $\sigma=0.05,
0.1$ and $\mu=\sigma
\left(\sigma-\sqrt{\sigma^2+4 r_+^2}\right)/(2
r_+)$. These initial wavepackets
correspond to perturbations with zero Aretakis
constant cases in eRN case \cite{Lucietti:2012b}.

\begin{figure}[!htbp]
\centering
\includegraphics[scale=0.51]{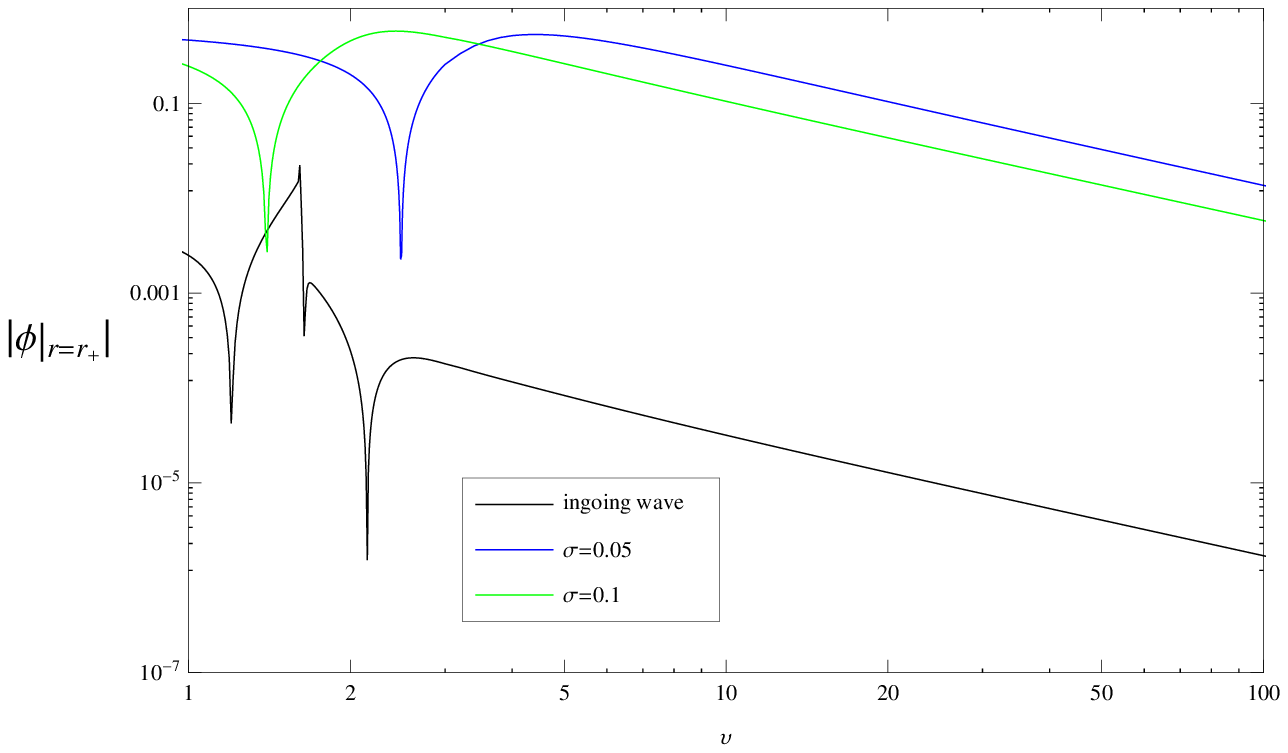}
\includegraphics[scale=0.55]{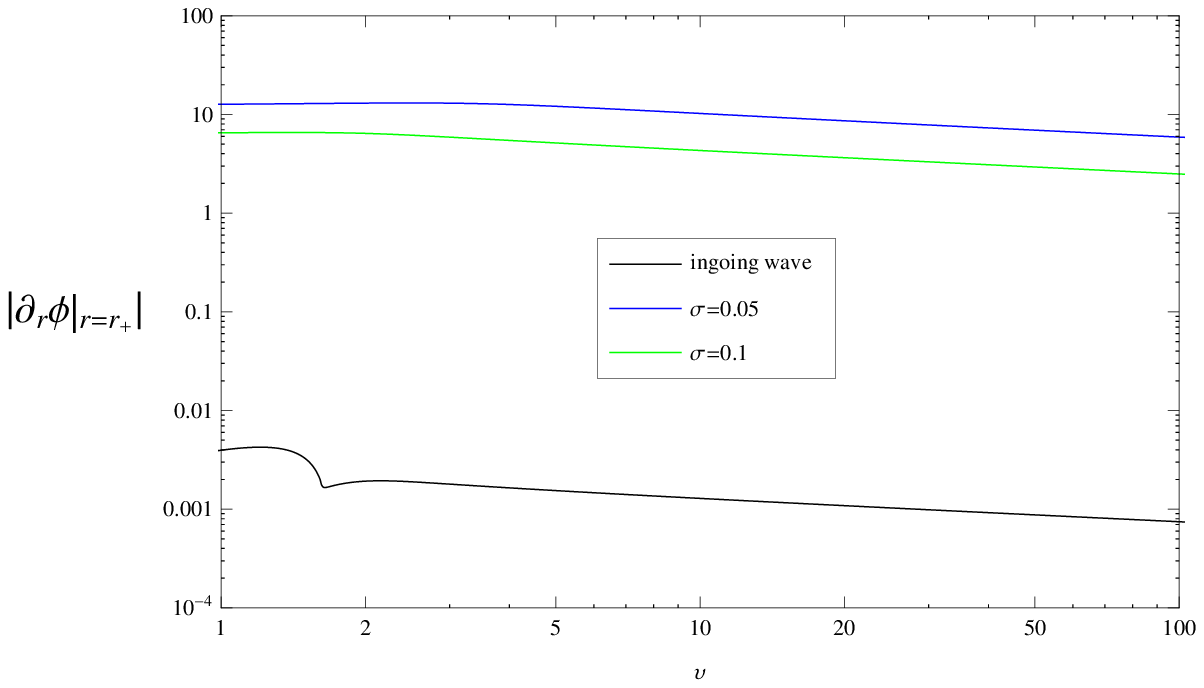}
\includegraphics[scale=0.53]{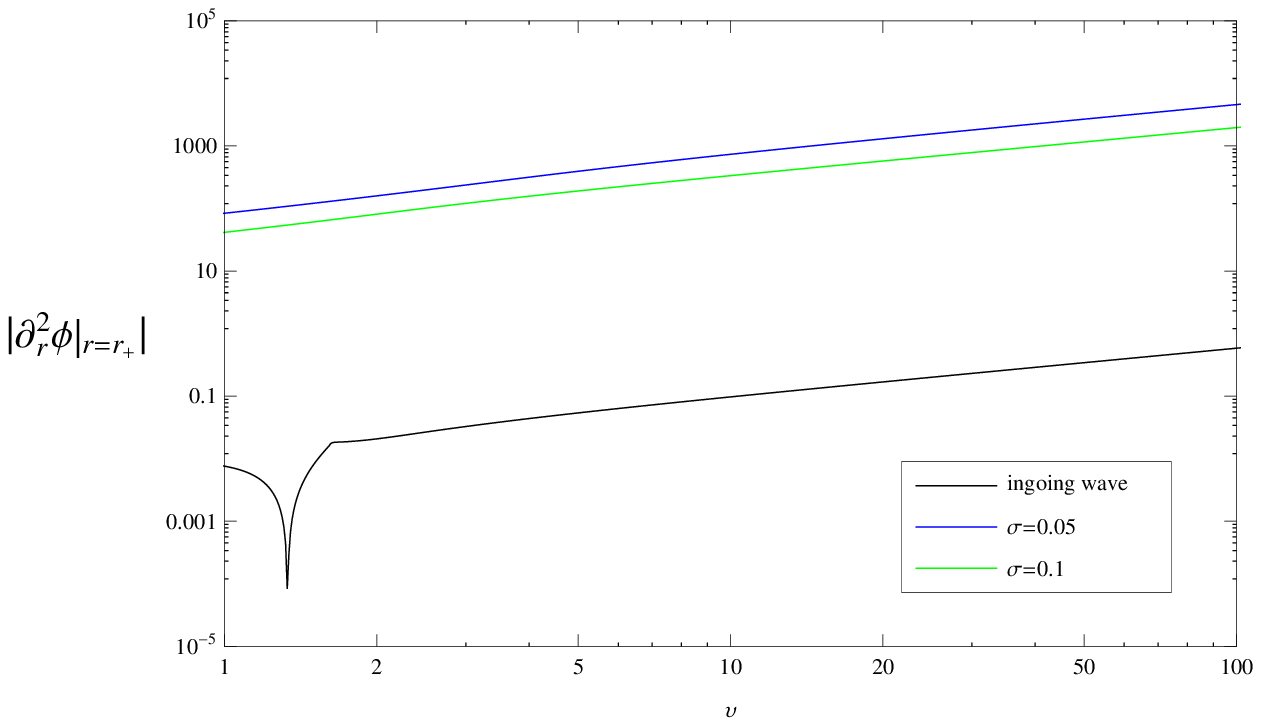}
\includegraphics[scale=0.55]{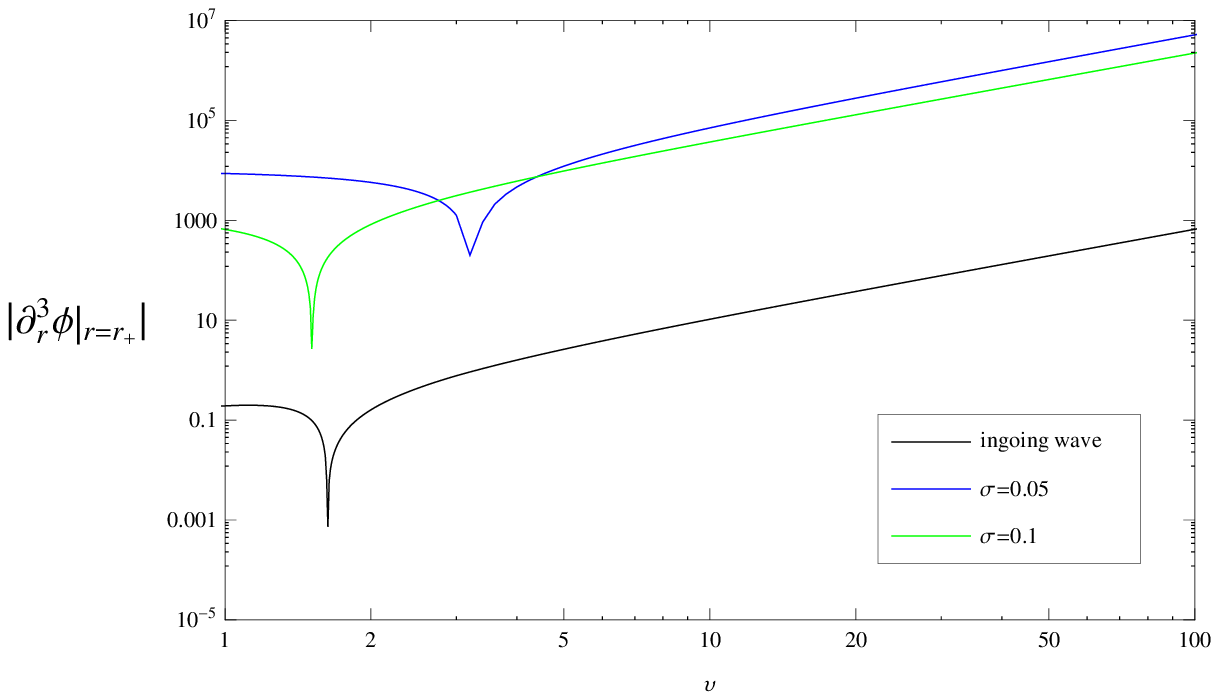}
\caption{Time evolution of $\phi |_{r=r_+},
\partial_r \phi |_{r=r_+}, \partial_r^2 \phi
|_{r=r_+}$ and $\partial_r^3 \phi |_{r=r_+}$ for
$l=1$. For the ingoing wavepacket, the initial
data takes $(\sigma', \mu')=(3.0, 10.0)$, and for
the outgoing wavepacket, the initial data takes
$\sigma=0.05, 0.1$ and $\mu=\sigma
\left(\sigma-\sqrt{\sigma^2+4 r_+^2}\right)/(2
r_+)$. We can see that: $\phi |_{r=r_+}$ and
$\partial_r \phi |_{r=r_+}$ exhibit a power-law
decay, while $\partial_r^2 \phi |_{r=r_+}$ and
$\partial_r^3 \phi |_{r=r_+}$ take a power law
blow-up.  This also implies that, for $l=1$ mode,
horizon instability starts to appear from the
second derivative of $\phi$.}
\end{figure}

From Fig. 6, we can see that $\phi |_{r=r_+}$ and
$\partial_r \phi |_{r=r_+}$ exhibit a power-law
decay, while $\partial_r^2 \phi |_{r=r_+}$ and
$\partial_r^3 \phi |_{r=r_+}$ take a power law blow-up.  This
implies that, for all cases we have considered
here, horizon instability appears starting from
the second derivative of $\phi$. This is
different from that in the eRN case
\cite{Lucietti:2012b}, in which horizon
instability starts to appear from the fourth
derivative of $\phi$.

Fitting  the late time behavior of $\phi
|_{r=r_+},\partial_r \phi |_{r=r_+}, \partial_r^2 \phi
|_{r=r_+},\partial_r^3 \phi |_{r=r_+}$ and $\partial_r^4
\phi |_{r=r_+}$ to $v^a$, we can find the
exponent $a$ for these functions. We list all the
fitting results in table I.

\begin{table}[!htbp]
\begin{center}
\begin{tabular}{c|cccccc}
\hline\hline
~~~&\multicolumn{3}{c}{Type I perturbations}&\multicolumn{3}{c}{Type II perturbations} \\
~~~$(\sigma, \mu)$&~~~~$(0.1, 0)$&~~~~$(0.1, -0.05)$&~~~~$(0.05,0)~~~~$&~~~~ingoing wave&~~~~$0.05$&~~~~$0.1$~~~\\
\hline
~~~$\phi |_{r=r_+}$ &~~~~$-1.191$ &~~~~$-1.230$ &~~~~$-1.127$~~~~ &~~~~$-1.246$ &~~~~$-1.252$ &~~~~$-1.243$~~~\\
\hline
~~~$\partial_r \phi|_{r=r_+}$ &~~~~$-0.218$ &~~~~$-0.229$ &~~~~$-0.206$~~~~ &~~~~$-0.230$ &~~~~$-0.233$ &~~~~$-0.234$~~~ \\
\hline
~~~$\partial_r^2 \phi|_{r=r_+}$ &~~~~$0.740$ &~~~~$0.757$ &~~~~$0.713$~~~~ &~~~~$0.768$ &~~~~$0.772$ &~~~~$0.763$~~~ \\
\hline
~~~$\partial_r^3 \phi|_{r=r_+}$ &~~~~$1.703$ &~~~~$1.746$ &~~~~$1.645$~~~~ &~~~~$1.770$ &~~~~$1.781$ &~~~~$1.762$~~~ \\
\hline
~~~$\partial_r^4 \phi|_{r=r_+}$ &~~~~$2.668$ &~~~~$2.745$ &~~~~$2.587$~~~~ &~~~~$2.772$ &~~~~$2.787$ &~~~~$2.769$~~~ \\
\hline\hline
\end{tabular}
\caption{Fitting results of the exponent $a$ for the function
$v^a$ with $l=1$. We fit the absolute values of $\phi$ and its
derivative up to the fourth level to $v^a$ for $80\leq v\leq 100$.
In type I perturbations, initial data are chosen to be $(\sigma,
\mu)=(0.1, 0), (0.1, -0.05), (0.05, 0)$. In type II perturbations,
parameters are chosen to be $(\sigma', \mu')=(3.0, 10.0)$ for the
ingoing wave, and $\sigma=0.05, 0.1$ and
$\mu=\sigma(\sigma-\sqrt{\sigma^2+4 r_+^2})/(2r_+)$ for the outgoing
wave. From the table, we can see that, for large $v$, the late time
behavior of $\partial_r^n \phi |_{r=r_+}$ is about $v^{a_0+n}$ with
$a_0 \sim -6/5$ and $a_0 \sim -5/4$ for type I and type II
perturbations, respectively.}
\end{center}
\end{table}

From table I, we can see that, for type II perturbations, the late time behavior of $\phi |_{r=r_+}$ is about
\begin{eqnarray}
\phi |_{r=r_+} \sim C v^{-5/4} \qquad v\rightarrow \infty \quad . \label{L1_outgoing-decay}
\end{eqnarray}

This is very different from eRN case
\cite{Lucietti:2012b}, where $\phi |_{r=r_+} \sim
v^{-3}$ as $v\rightarrow \infty$. It shows that
in the eRN-AdS black hole background, the
massless scalar perturbation decays much slower
compared with the eRN black hole case. Moreover,
we can see from the table that the late time
behavior of $\partial_r^n \phi|_{r=r_+}$ takes a
power-law as $v^{a_0 + n}$ with $a_0 \sim -6/5$
and $a_0 \sim -5/4$ for type I and type II
perturbations, respectively.  This is consistent
with analytical result (\ref{latetime}). The blow
up appears earlier and more violent than that in
the eRN case \cite{Lucietti:2012b}, where
$\partial_r^4 \phi|_{r=r_+}$ approaches to
$v^{2.7} (v^{2.8})$  in eRN-AdS while $v^2 (v)$
in eRN case for type I (II) perturbations.

Comparing with $l=0$ mode discussed in last
section, the late time behavior of $\phi
|_{r=r_+}$ with $l=1$ for ingoing wave initial
data takes a moderate power-law decay. This is
contrary to the observation in the eRN case
\cite{Lucietti:2012b}, but is consistent with
results in the nonextreme RN-AdS black hole
\cite{Zhu:2001a,Wang:2004a}.

\subsubsection{For $l>2$ modes}

We also extend our numerical calculation to the
$l=2,3,4$ and $5$ modes to see further the effect
of $l$ on the decay law of $\phi |_{r=r_+}$ and
on the blow-up behavior of $\partial_r^n \phi
|_{r=r_+} (n>0)$. We consider both the outgoing
wave initial data and an ingoing wave initial
data. We observe that the late time behavior of
$\phi |_{r=r_+}$ for these modes are all
power-law decay. This further confirms the
argument in \cite{Wang:2004a} that when the
nonextreme AdS becomes extreme, the exponential
decay of the perturbation will give way to the
power-law decay. Also by fitting values of $\phi
|_{r=r_+}$ to $v^a$, we can find the exponent $a$
for all these modes, which we list in table II.
We plot the results of the late time tails of
ingoing wave case in fig. 7.

\begin{table}[!htbp]
\begin{center}
\begin{tabular}{c|cccccc}
\hline\hline
~~~&~~~~$l=0$&~~~~$l=1$&~~~~$l=2$&~~~~$l=3$&~~~~$l=4$&~~~~$l=5$~~~\\
\hline
~~~outgoing wave &~~~~$-1$ &~~~~$-1.230$ &~~~~$-1.563$ &~~~~$-1.865$ &~~~~$-2.286$ &~~~~$-2.631$~~~\\
\hline
~~~ingoing wave &~~~~$-2$ &~~~~$-1.246$ &~~~~$-1.557$ &~~~~$-1.926$ &~~~~$-2.238$ &~~~~$-2.605$~~~ \\
\hline\hline
\end{tabular}
\caption{Fitting results of the exponent $a$ for the function
$v^a$ for different modes. We fit the absolute values of $\phi |_{r=r_+}$ to $v^a$ for $80\leq v\leq 100$.
The parameters of the initial perturbations are chosen to be $(\sigma,
\mu)=(0.1, -0.05)$ and $(\sigma', \mu')=(3.0, 10.0)$ for the outgoing and ingoing wave, respectively.}
\end{center}
\end{table}

\begin{figure}[!htbp]
\centering
\includegraphics[scale=0.6]{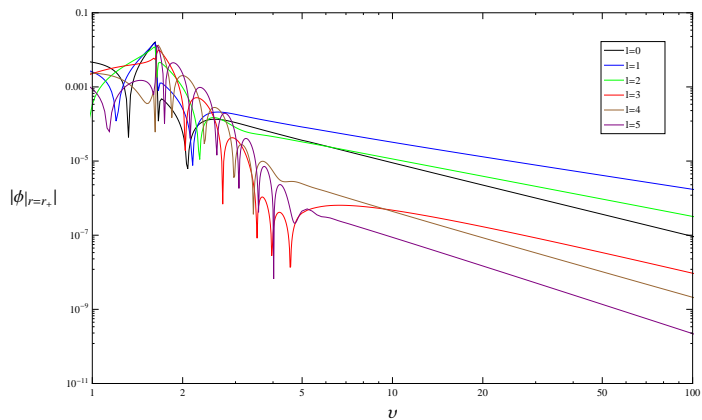}
\caption{Time evolution of $\phi |_{r=r_+}$ for different $l$ for the ingoing wave. Parameters are chosen to be
$(\sigma', \mu')=(3.0, 10.0)$. All these modes at late time are power-law decay. And modes with $l\geq 4$ decay faster than the $l=0$ mode.}
\end{figure}

From table II and fig. 7, we can see that at late
time: (1) For the outgoing wave with non-zero
Aretakis constant $H_0$, mode with higher $l$
will decay faster; (2) For the ingoing wave (with zero Aretakis constant $H_0$), when $l>0$,
the first three modes with $l=1,2,3$  decay
slower than the fundamental mode; while when
$l\geq4$ it decays faster than the $l=0$ mode.
This implies that the decay of the massless
scalar field will be dominated by the first few
lower-$l$ modes. This is in some similarity with
the situation in de-Sitter black hole, where the
late time behavior of massless scalar ingoing
perturbations is \cite{Brady:1996a,Brady:1999a}
\begin{eqnarray}
|\phi| \sim \left\{\begin{array}{ll} e^{-l \kappa_c t} & l>0\\
|\phi_0|+|\phi_1| e^{-2\kappa_c t} & l=0
\end{array}\right., \label{dSBH}
\end{eqnarray}
The item $\kappa_c$ is the surface gravity on the
cosmological horizon and $\phi_0, \phi_1$ are
some constants. In the de Sitter case, the decay
law of different modes are also divided into two
branches, $l=0$ mode and $l>0$ modes, and for
$l>2$ it decays faster than the fundamental mode.

From table I and II, we can also see that, for $l>0$
modes, horizon instability starts to appear from
the second ($l=1,2,3$) or third ($l=4,5$)
derivative of $\phi$, earlier than that in the
eRN case.

\section{Conclusions and Discussion}

Recent studies have proved that a massless scalar field has an instability at the horizon of an
extreme Reissner-Nordstr\"om black hole. Considering that a scalar field will confront
different boundary conditions when it propagates in the AdS background, we have extended the
stability study to extreme RN-AdS black hole. We have studied the massless scalar perturbations
for different angular index. For $l=0$ mode, we can define the Aretakis constant $H_0$, and hence
have shown the horizon instability analytically by assuming that the scalar field $\phi
|_{r=r_+}$ on the horizon decays at late time.

Furthermore, we have applied numerical
calculation and found the supporting evidence of
the decay of $\phi |_{r=r_+}$.  We have extended
our numerical computation to the higher modes and
found the consistent power-law decay of the
massless scalar field perturbation. The decay of
the scalar field is not exponential as observed
in the nonextreme AdS black
hole\cite{Horowitz:1999a,Horowitz:1999b,Wang:2000b}.
In \cite{Wang:2004a} it was argued that when the
nonextreme AdS black hole approaches to extreme
hole, the exponential decay will give way to the
power-law decay. Our numerical result for the
eRN-AdS black hole have supported this argument.
This is an important point, since it may lead to
a departure from stability already at the
nonextreme level. While we expect some kind of
instability in the extreme limit, already from
the very fact that the extreme limit corresponds
to a zero temperature thermodynamics, it is
worthwhile checking whether the stability has
stronger roots. At this point, a nonlinear
approximation, or at least a backreaction
calculation should be important for the
clarification of the stability or instability
determination. Indeed, in case the backreaction
pulls the black hole out of the extreme limit,
there are two choices, namely either the
nonextreme limit is stable, thus backreaction
stabilizes the problem, or the quasiextreme case
is also unstable, and the whole black hole is
unstable. Very recently, there appears a work in
this direction for the eRN case
\cite{Murata:2013a}. It was found that
generically the endpoint will be a stationary
nonextreme black hole, but if there exists
non-generic initial perturbations,  the
instability will never end. We hope that it can
be extended to eRN-AdS case in near future.

The dependence of the late time tail on the
angular index shown in the eRN-AdS black hole
here is different from that in the eRN black
hole, which reflects the influence of the
spacetime on the perturbation at late time.

In the eRN-AdS black hole, we have found that
when $l=0$ the horizon instability starts to
appear from the second or third derivative of the
scalar field when the Aretakis constant is
nonzero or zero.  When $l>0$, the horizon
instability starts from the second ($l=1,2,3$) or
third ($l=4,5$) derivative of the scalar field.
This is different from that in the eRN case,
where the blow-up appears at higher derivative of
the scalar field, especially for higher angular
index case \cite{Lucietti:2012b}. This shows that
the instability in the extreme AdS black hole can
happen more easily than that in the extreme
asymptotically flat black hole.

\section*{Acknowledgments}

This work is partially supported by the National
Natural Science Foundation of China and FAPESP in
Brazil.

\end{document}